\documentclass[vecphys]{svmult}
\usepackage{makeidx}
\usepackage{graphicx}
\usepackage{multicol}

\def\gsim{\mathrel{\raise.5ex\hbox{$>$}\mkern-14mu
             \lower0.6ex\hbox{$\sim$}}}

\def\lsim{\mathrel{\raise.3ex\hbox{$<$}\mkern-14mu
             \lower0.6ex\hbox{$\sim$}}}

\makeindex

\begin{document}

\title*{Cosmological Inflation: A Personal Perspective}

\titlerunning{Cosmological Inflation}

\author{D. Kazanas}

\institute{Astrophysics Science Division, Code 663, NASA/GSFC,
Greenbelt, MD 20771 \texttt{Demos.Kazanas@nasa.gov}}

\maketitle

\begin{abstract}

We present a brief review of Cosmological Inflation from the
personal perspective of the author who almost 30 years ago proposed
a way of resolving the problem of Cosmological
Horizon by employing certain notions and developments from the field of
High Energy Physics. Along with a brief introduction of the Horizon
and Flatness problems of standard cosmology, this lecture
concentrates on personal reminiscing of the notions and ideas that
prevailed and influenced the author's thinking at the time.
The lecture then touches upon some
more recent developments related to the subject and concludes with
some personal views concerning the direction that the cosmology
field has taken in the past couple of decades and certain
speculations some notions that  may indicate future directions of
research.

\end{abstract}

\section{Introduction}

The development of General Relativity and the possibility it offers
to probe the issues of the overall geometry, topology and evolution
of the Universe as a whole it is certainly one of the great
achievements of human spirit and captured my own imagination
when I first came across an article by G. Chasapis on the
``Universe" in the Greek encyclopedia ``Helios". Since then, the
field of Cosmology has been for me an avocation of sorts, honed in
time,  as I pursued studies in physics first at the University of
Thessaloniki, through courses offered by professors G. Contopoulos
and S. Persides and then through a large number of discussions with
my late roommate and fellow graduate student at the University of
Chicago B. Xanthopoulos as well as from interactions with my late
thesis advisor D. N. Schramm. Since this is a brief personal account
and not a review, I would like to apologize in advance to many for
the absence of a large number of important references and contributions
to the subject.

Following the original cosmological models of
Einstein, de Sitter, Lemaitre, Friedman and others and the discovery
of the expansion of the Universe by Hubble, the next development
came through the realization (Gamow, Alpher) that the present
expansion of the Universe implies that at an earlier stage it should
have been sufficiently hot for nuclear reactions to take place. This
then, supported by the discovery of the Cosmic Microwave Background
(CMB) radiation, led to the development of Big Bang Nucleosynthesis
(BBN) by Wagoner, Fowler \& Hoyle \cite{BBN}  that still
serves as a ruler against which all cosmological models have to be
measured.

In the mid to late 70's, the subject of Cosmology was much less
prominent than today, at least from the perspective of a
graduate student, even one that specialized in astrophysics.
The primary Cosmology text was Weinberg's book
\cite{W72}, wherein one could find the fundamentals of General
Relativity and its application to relativistic objects, i.e. neutron stars
and black holes, as well as the Universe itself. Its exposition of Cosmology
provided, in addition to the general cosmological models, also the
details of the thermal evolution of the universe and some of the
open outstanding issues of standard cosmology namely the entropy (number
of photons) per baryon $1/\eta$ in the
cosmological fluid, and the issue of horizons.

The issue of the high value of $1/\eta~  (\simeq 10^9)$,
compared to that found in a typical star ($\eta \simeq 1$), was given
a prominent position both in \cite{W72} and also in Weinberg's, then
new, more popular book ``The First Three Minutes"\cite{W3min}. Particular
emphasis was given at the difficulty of producing such a large value
for $1/\eta$ through dissipative processes given that the homogeneity
and isotropy of the Universe that allows only for the effects bulk
viscosity. However, as argued by the author, even this process could
not add much more than a photon per baryon to the value of $1/\eta$.

\section{The Cosmological Problems}

At this point I would like to make a brief digression to outline the
dynamics of the Universe and formulate the Cosmological problems of
Horizon and Flatness. In my view, the root of both these problems,
at least partially, lies in the fact that, in the system of units in
which $h = c = 1$, the gravitational constant $G$ has dimension of
(mass)$^{-2}$, the so-called Planck mass; this is the mass of
particles for which the Schwarzschild and Compton lengths are equal,
i.e. $2 G M_P/c^2 = h/M_P c$, or $G = hc/2 M_P^2$ or $M_P =
(hc/G)^{1/2} \simeq 10^{-5}$ gr. To this mass scale one can assign
equivalent length, time and temperature scales of corresponding
values $l_P \simeq 10^{-33}$ cm, $t_P \simeq l_P/c \simeq 10^{-43}$
sec and $T_P \simeq 10^{32}$ K.

\subsection{Newtonian Cosmology}

It is most amazing that the dynamics of the Universe as determined
by the equations of General Relativity can be derived from purely
Newtonian considerations. The facts that allow a Newtonian treatment
of cosmology are that: (1) the Universe is homogeneous and
isotropic, so any point can serve as the origin of a spherically
symmetric coordinate system and (2) the property of the Newtonian
potential that for a spherically symmetric matter distribution, the
dynamics of the matter within a volume of radius $a$ is determined
only by the matter interior to $a$. Therefore, for a homogeneous and
isotropic distribution, such as that of the Universe, one can choose
the radius $a$ arbitrarily and study the dynamics this sphere, all
matter exterior to $a$ being irrelevant. The Hubble law indicating
that velocities are proportional to the distance, then, guarantees
that shells of different radii expand homologously and do not run
onto each other.

One can, hence, write the equations of motions of a sphere of
arbitrary radius $a$ simply using the total energy
integral, $E$, namely
\begin{equation}
\frac{1}{2} \dot a^2 - \frac{G M}{a}=\frac{1}{2} \dot a^2 -
\frac{4\pi G \rho}{3}a^2= E ~~~{\rm or} ~~~ H^2 = \frac{\dot
a^2}{a^2} = \frac{2E}{a^2} + \frac{8\pi G \rho}{3} \label{MM}
\end{equation}
It is instructive to compare this equation the the corresponding
Einstein equation for a homogeneous and isotropic Universe of
spacial curvature $k = 1, ~0,-1$ corresponding to a closed, flat or
open Universe:
\begin{equation}
\frac{\dot a^2}{a^2} + \frac{k}{a^2} = \frac{8\pi G \rho}{3}
\label{EE}
\end{equation}
The role of the energy is played by the spatial curvature, $-k$, indicating
that in a closed Universe ($k>0, ~E<0$) the radius of the sphere reaches a
maximum while in flat and open universes it can reach infinity.

The solution of this equation requires an assumption about the
variation of the density with time (or with $a$); this can be
obtained from the conservation of energy, which reads $\rho \propto
a^{-3}$ for pressureless matter and $\rho \propto a^{-4}$ for radiation while
for temperature implies $T \propto 1/a$.

The only difference between the Newtonian and Einstein version of
Cosmology becomes apparent only by differentiating Equations
(\ref{MM}) or ({\ref{EE}) taking into account the relation between
$\rho$ and the pressure $P$ from local energy conservation (Eq.
\ref{EC} below) to obtain the corresponding force equation
\begin{equation}
\frac{\ddot a}{a} = - \frac{4 \pi G}{3}(3P + \rho)~.
\end{equation}
This equation incorporates the contribution of pressure to the
gravitational force, as it should, since pressure is energy density
and all energy gravitates. The presence of this term, significant in
the radiation era, has been verified by comparing the outcome of BBN
to observation\cite{BBN08}.

\subsection{The Horizon Problem}

The finite age of the Universe $t_U \simeq 1.4 \times 10^{10} \;{\rm
yr} \simeq 5 \times 10^{17}$ sec, along with the finite speed of
light indicate that light signals since the creation of the Universe
have traveled a distance $R_H \simeq c t_U \simeq 10^{28}$ cm. One can
now estimate the size of $R_H$ at the time its age was  $t_P$ and
its temperature $T_P$, by scaling $R_H$ by the ratio of the CMB
temperatures at the two epochs, namely $R_P \simeq R_H (3 {\rm K}/10^{32}~
{\rm K}) \simeq 10^{-3}$ cm.  This size is 30 orders of magnitude larger
than the horizon size at that time $c t_P \simeq l_P$, indicating that
the Universe at that time comprised $\sim 10^{90}$ causally disconnected regions, all
of which must have had approximately the same temperature since the
Cosmic Microwave Background (CMB) appears to be quite uniform across
the observed Universe. This constitutes the Horizon problem.
More formally, the size of the horizon must take into account the
fact that the photon signal co-moves with the expanding Universe and
it is thus given by
\begin{equation}
S_H = a(t) \int_0^t \frac{c dt}{a(t)}
\end{equation}
One can see that for a power-law expansion rate $a(t) \propto t^p$
with $p<1$ (as is the case for a radiation ($p = 1/2$) or  matter
($p = 2/3$) dominated Universe) the horizon size is just a multiple
of $c t_U$. However, for $p \ge 1$ the integral is dominated by the lower
limit and the horizon diverges at $t \rightarrow 0$.

\subsection{The Flatness Problem}

For a given value of the ratio $H \equiv \dot a/a$, Equation
(\ref{MM}) defines a characteristic value of the density $\rho_c =
3H^2/8 \pi G$,  i.e. the density for which the explosion energy $E$
is equal to zero, and use it to define the ratio of the density to the
critical one as $\Omega = \rho/\rho_c$. We can now divide Eq.
(\ref{MM}) by $\dot a^2$ to obtain
\begin{equation}
1- \Omega = \frac{2E }{\dot a^2} =  - \frac{k}{(Ha)^2}
\label{flat}
\end{equation}
Applying the above relation at two different values of $a$ and the
corresponding values of $\Omega$ we obtain
\begin{equation}
\Omega_1 - 1 = \frac{\dot a_0^2}{\dot a_1^2}(\Omega_0 -1)
\end{equation}
One can now see that if the present value $\vert \Omega_0-1\vert
\simeq {\cal O}(1) $, then, given than in standard cosmology $a(t)
\simeq K \, t^{1/2}$, $\dot a_0^2/\dot a_1^2 \simeq t_1/t_0$; since
$t_0 \simeq 10^{17}$ sec, at an earlier epoch with $t_1 \ll
t_0$, $\Omega_1 \rightarrow 1$. If, in particular we set $t_1 \sim
t_P \sim 10^{-43}$ sec, $t_1/t_0 \simeq 10^{-60}$, i.e.
under Standard Cosmology, at the Planck time, the radiation density was
equal to the critical denisty to within 1 part in $10^{60}$!

\section{Phase Transitions, Baryogenesis}

The focus placed by Weinberg on the value of $\eta$
helped galvanized a couple of fellow graduate students including
myself to take an independent look at this parameter in search for
mechanisms that could account for its value. As far as I
can now recall, our first attempt was to use Weinberg's prescription of
bulk viscosity\cite{Wbulk} but dare to consider its application to
much higher temperatures and include much more
massive particles than had been considered till then. However, we soon
realized that no matter what the temperature and the particle masses,
this process could add but a small number of photons
per baryon in the cosmological fluid.

In search of other entropy producing processes I stumbled upon the
idea of phase transitions and the entropy associated with the
latent heat. Being aware that quarks were confined
into baryons by a potential that grows (linearly) with distance, I
considered that if this transition could be somehow delayed during
the expansion of the Universe to densities lower than nuclear, the
linear quark interaction could produce extremely large values of
entropy {\em from the vacuum!} Because I considered such a situation
rather contrived and poorly constrained, I suggested (in a
publication\cite{Kaz78} that received just a single
citation\cite{DZ}) that, even though there are overall no free
quarks, it is possible that within a horizon volume  there may be an
excess of color, which would now interact via the quark linear
potential with a similar color excess in an adjacent horizon volume.
Assuming that the local color excess to be purely statistical, i.e.
proportional to the square root of the particles within a given
horizon volume, then one can calculate the amount of entropy
produced as the universe expands. However, under these conditions
the entropy thus released does not contribute significantly to
$1/\eta$. Despite this fact, I was impressed by the possibility of
energy production from the vacuum and thought it could have
potentially significant consequences.

The issue of the value of $1/\eta$ was resolved in 1978 in an
altogether different and far more subtle way (e.g. \cite{Yo78, DS}
and others): The production of a large number of photons per baryon
was supplanted by the production of a small excess of baryons over
antibaryons in an originally symmetric cosmic fluid; this entailed
invoking processes that violated baryon conservation, the $CP$
symmetry and thermodynamic equilibrium. These processes were
apparently possible within the context of Grand Unified Theories,
i.e. theories that unified the strong with the weak and
electromagnetic interactions at energies $\sim 10^{15}$ GeV.

While the issue of the photon to baryon ratio $1/\eta$ was resolved
in principle as above, the issue of entropy production from the
vacuum was still extremely appealing to me and my thought was that
perhaps this could help resolve the remaining open cosmological
problem, that of the Horizon.

\section{Resolving the Horizon Problem}

At the end of 1978 I got my PhD, left Chicago and spent the
following year (1979) in the Greek military. Upon my discharge I
returned to the US having been offered an NRC fellowship at GSFC by
Floyd Stecker. On my way back to the US I spent a few days at
Nordita in Denmark, where K. Sato had been also a visitor. He was
very much interested in phase transitions in the early universe and
we did discuss some of the issues of the quark - baryon one outlined
above with one of his comments being that he was interested ``in a
different type of phase transition".

This last comment caught my attention enough to launch a (not so
thorough, as it turned out) search for this different type of phase
transition; the search produced only one relevant paper
\cite{Lash79}, which however involved the quark-baryon transition I
was already aware of. At the same time, my interest in the horizon
problem was rekindled by a paper by Brown \& Stecker\cite{BS79},
which considered the intriguing and interesting possibility of a
matter - antimatter domain Universe produced by a phase
transition-like violation of the CP symmetry with the order
parameter taking randomly values of either $-1$ or $+1$ within each
domain. The Horizon Problem is at the very heart of this proposal
because the size of these domains is limited by the Horizon size at
temperatures $\sim 10^{15}$ GeV, at which the baryon asymmetry is
formed. In one of the references of \cite{BS79} I found then a
citation to \cite{ZKO} who discussed very much the same problem. The
authors of \cite{ZKO} showed that because of the discrete nature of
the CP--symmetry the corresponding phase transition produced a
network of walls separating the two phases and that the wall network
corresponds to a perfect fluid with equation of state $P = -2 \rho
/3$; this then leads to an expansion rate for the Universe $a(t)
\propto t^2$ which, as discussed above can lead to domain sizes
sufficiently large to avoid contradiction with observations on the
existence of antimatter in space. This provided a resolution of
sorts of the Horizon Problem, except for the fact that the resulting
Universe would be very inhomogeneous due to the presence of these
walls, in contradiction with observation.

At this point, I noticed a paper \cite{Zee} discussing phase
transitions within the context of Spontaneous Symmetry Breaking
(SSB), a subject that had been extensively treated by \cite{KL76,
L79}. These are not unlike those discussed in \cite{ZKO} but the
broken symmetries are not necessarily discrete and hence they do not
have to lead to inhomogeneities. The work of \cite{KL76} was very
instructive: It showed that the energy stored in the
self-interacting Higgs field $\phi$, a fundamental ingredient of SSB,
acts as a perfect fluid with an equation of state $P_v = -\rho_v$
with the vacuum expectation value of $\phi$ and the energy density
$\rho_v$ having the temperature dependence given in Fig. 1: ($i$)
For $T>T_c$, $\langle \phi \rangle =0$ and its energy density is
$\rho_v \simeq a_{bb}T_c^4 = {\rm constant} <  \rho_r = a_{bb}
T^4$ ($\rho_r$ is the radiation energy density, $a_{bb}$ is the black
body constant). ($ii$) For $T < T_c$, $\langle \phi \rangle \ne 0$ and the energy
density $\rho_v = \epsilon_v = a_{bb}T^4$ decreases with
the temperature $T$ but remains comparable to that of radiation $\rho_r$.
This phase transition does not involve the confinement of quarks and
does not suffer from the problems with that discussed earlier.

\begin{figure}
 \vspace*{-2mm}
\centering
\includegraphics[width=8cm]{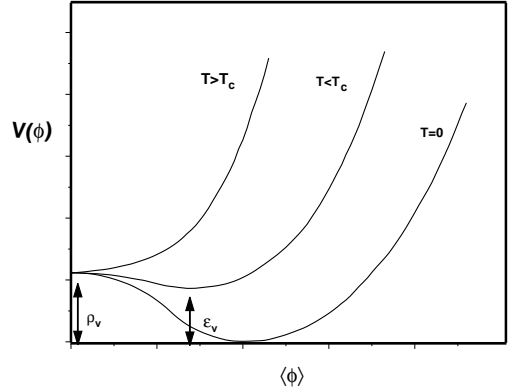}
%
%
\vspace*{-2mm} \caption{\small The temperature dependent Higgs potential.
For $T>T_c$, $\left< \phi \right> =0$ and the vacuum energy density
$\rho_v$ is constant but insignificant. For $T < T_c$ $\left< \phi
\right> \ne 0$ and the vacuum energy density $\epsilon_v$ depends on
$T$, being zero for $T \rightarrow 0$. }
\label{Lag1}       
\end{figure}

The effects of such a phase transition on the evolution of the
Universe can be easily studied  by considering that the total
pressure and energy density consist of the sum of radiation and the
vacuum, i.e. $P = P_r + P_v$ and $\rho = \rho_r + \rho_v$, with each
obeying its own equation of state, i.e. $P_r = \rho_r/3$ and $P_v =
-\rho_v$.  The solution to Einstein's equation
\begin{equation}
\left(\frac{\dot a}{a} \right)^2 + \frac{k}{a^2} = \frac{8 \pi G}{3}
(\rho_r + \epsilon_v)
\end{equation}
requires also the knowledge of variation of $\rho, ~\epsilon_v$ with time
or with $a$. The relation between $a$ and $T$ is given by the first
law of thermodynamics \cite{Kaz80}
\begin{equation}
a^3 \frac{d \rho}{da} = - 3(P + \rho)a^2  ~~~{\rm or}~~~a^3 \frac{d
}{da}(\rho_r + \epsilon_v) = - 4 \rho_r a^2 \label{EC}
\end{equation}
after taking into consideration the corresponding equations of
state. This last equation then leads to the following relations
between $a$ and $T$ or $\epsilon$ and $T$:
\begin{equation}
a \propto \frac{1}{T^2} ~~~~{\rm or} ~~~~ T \propto \frac{1}{a^{1/2}} ~~~~~
{\rm and} ~~~~~ \epsilon_v, \; \rho_r \propto
\frac{1}{a^2} \label{tvsa}
\end{equation}

The presence of a vacuum component makes therefore a great deal of
difference in the cooling of the universe: As long as $\epsilon_v
\propto T^4$, the universe has to expand by twice as many decades
to cool by the same factor as under adiabatic conditions. During
this period the vacuum energy is never dominant but it is comparable
to that of radiation and keeps feeding into it as $\epsilon_v$
slowly decreases. It was argued in \cite{Kaz80} that this behavior
should terminate at some point, else it would over-dilute the
baryon/photon ratio.

With the relation between $a$ and $T$ (Eq. \ref{tvsa}) it is easy to
compute the evolution of $a$ (Eq. \ref{EE}) to obtain $a \propto t$,
indicating that the horizon diverges logarithmically for $t
\rightarrow 0$. However, this divergence is very mild and it is
unlikely that it can resolve the Horizon Problem. Motivated by the
work of \cite{KL76, L79} and prompted by the referee of the paper I
had submitted I considered also the case $\epsilon_v \propto T^2$
for $T < T_c$. The slower decrease in the vacuum energy density then
gave a very different relation between $a$ and $T$,
namely\cite{Kaz80}
\begin{equation}
\frac{a}{a_c} = \frac{T_c}{T} exp \left[ \frac{1}{4}
\left(\frac{T_c^2}{T^2}-1 \right) \right] ~~~~{\rm for} ~~T < T_c
\end{equation}
where $a_c$ is the value of $a$ when the temperature drops to the
critical one $T_c$.  This expression leads to a much slower decrease
of $T$ with $a$, which, when substituted into Eq. (\ref{EE}) yields
an exponential expansion $a \propto exp[t^{1/3}]$, which can expand
the Horizon size to values much larger than $R_H$, thereby resolving
the Horizon Problem in a robust way. One can also see that an exponential
expansion quickly renders the RHS of Eq. (\ref{flat}) $\ll 1$, resolving
also the Flatness Problem.

\section{``Nothing Succeeds like Success"}

Considerations and calculations similar in spirit to those discussed
above were worked out at approximately the same time by Sato
\cite{Sato} and Guth \cite{Guth}; the early stage exponential
expansion of the Universe driven by the energy density of the vacuum
was given the name\cite{Guth} `Inflation', a term resonant with the
state of the US economy at the time, which has been since adopted
universally, despite the subsequent change in the state of the US
economy. The evolution of the Universe as described in \cite{Sato,
Guth} proceeds through the formation of bubbles with $\rho_v = 0$
surrounded by exponentially expanding space of $\rho_v \ne 0$; the
hope was that eventually the $\rho_v = 0$ regions would occupy the
entire volume of the universe, which in the mean time had inflated
enough to resolve the Horizon and Flatness problems. The problem was
that, due to a secondary minimum of $V(\phi)$ at $\phi=0$, the
transition rate to $\rho_v = 0$ was too slow to complete the
transition. This shortcoming was overcome in the `New Inflation'
\cite{AS} where the Universe was considered to `slowly roll' down on
a potential similar to that corresponding to $T=0$ in Fig. 1, with
the expansion dominated by a roughly constant $\epsilon_v$ and with
the present horizon constituting a small patch of the expanding
universe with $\Omega =1$ with extremely high accuracy.

However, the most important feature of the `New Inflation' is that
it affords a process that can produce the fluctuations necessary for
the formation of cosmological structure: During the `slow-roll'
period of the evolution of the Universe the geometry of space is
that of de Sitter space with a cosmological horizon at a constant
coordinate distance. Quantum fluctuations of the field $\phi$
created with constant amplitude $\delta \phi$ decrease until they
cross the de Sitter horizon; then, as they are stretched by the
expansion of the Universe to super-horizon scales, their amplitude
freezes to the value they had at horizon crossing; this is due to an
interplay between the scalar field and metric perturbations; in fact
because the field perturbation is proportional to $\delta \phi
\propto V_{,\phi}/V$ it increases toward the end of inflationary
phase. After the end of the phase transition, the Universe resumes
its conventional expansion; as the horizon size increases the
fluctuations come within the horizon at roughly the constant
amplitude they had when exiting the de Sitter horizon to produce the
Harrison--Zeldovich spectrum of cosmological perturbations. These
have subsequently left their imprint as fluctuations on the CMB
temperature which were recently measured by both the COBE and WMAP
\cite{Sper} missions confirming the general predictions of the
inflationary scenario.

A most interesting feature of the above process is that the
amplitude of perturbations depends on the shape of the potential
$V(\phi)$ and the energy scale of inflation.
Furthermore, small deviations of the fluctuation spectrum from the
precise Harrison--Zeldovich form, can also give an estimate of the
number of e--foldings of inflation which was found (for the simplest
models) to be of order of 60-70 (while it could, in principle, be
much larger) \cite{Sper}, suggesting an expansion by a factor of
roughly $10^{30}$, the minimum required to reconcile the disparity
between the size of the universe and the Planck length at $t=t_P$
discussed in \S 2.

While the issue of the horizon size or the flatness of the universe
are resolved in an appealing way by the inflationary scenario, these
issues provide little additional quantitative evidence in support of
its fundamental premises. However, the production, amplitude and
spectrum of the resulting matter fluctuations and their imprint on
the CMB, the result of the quantum fluctuations of the field $\phi$,
provides a unique to date method for the production of the
fluctuations necessary to produce the observed structure in the
Universe and a much more rigorous instrument of scrutiny of the
above ideas. The interested reader can find of all these in the
modern literature (e.g. the monograph by Mukhanov\cite{M05}). While
there has been at least one (sound in my opinion) objection against
the entire `Inflationary' edifice\cite{Penrose1}, in the absence of
a successful accompanying account of the CMB fluctuations, this has
gained little traction. Despite these objections and those raised in
the next section concerning the nature of the ingredients of the
Inflationary Paradigm, the success of this scheme in addressing the
CMB fluctuations make it an indispensable tool in modern cosmology;
it is then not unreasonable to conclude that in science as in
business ``nothing succeeds like success".

\section{Discussion and Speculations}

It is fair to say that the ideas of Cosmological Inflation provided
the impetus and the physical notions for tracing (with great
success) the evolution of the Universe to an era impossible to
imagine 30 years ago. In my personal view, a great deal of the
appeal of this scenario lies in its simplicity: The mathematics of
the original inflationary proposal are almost trivial, while the
complexity of even the theory of fluctuations is moderate.

That being said, again in my personal view, the characterization of
Inflation as a `scenario' rather than a `theory' is also not unfair.
To begin with, while in its original versions the scalar field
employed to resolve the cosmological puzzles was considered to be
the Higgs field, for reasons unknown to me, this association was
dropped in favor of an altogether independent scalar field $\phi$
(the inflaton), unencumbered by such an association (perhaps because
of the constraints it imposed on the models). Furthermore, the all
important self-interaction potential $V(\phi)$ of this field remains
(again in my personal, poorly informed view) in the realm of
phenomenology. Despite these objections, the concordance of its
predictions with the CMB data has set the bar for future
alternative, competing schemes.

\renewcommand{\thefigure}{2}
\begin{figure}[t]
\vspace*{-0.3cm}
\centerline{\hbox{\includegraphics[width=0.68\textwidth]{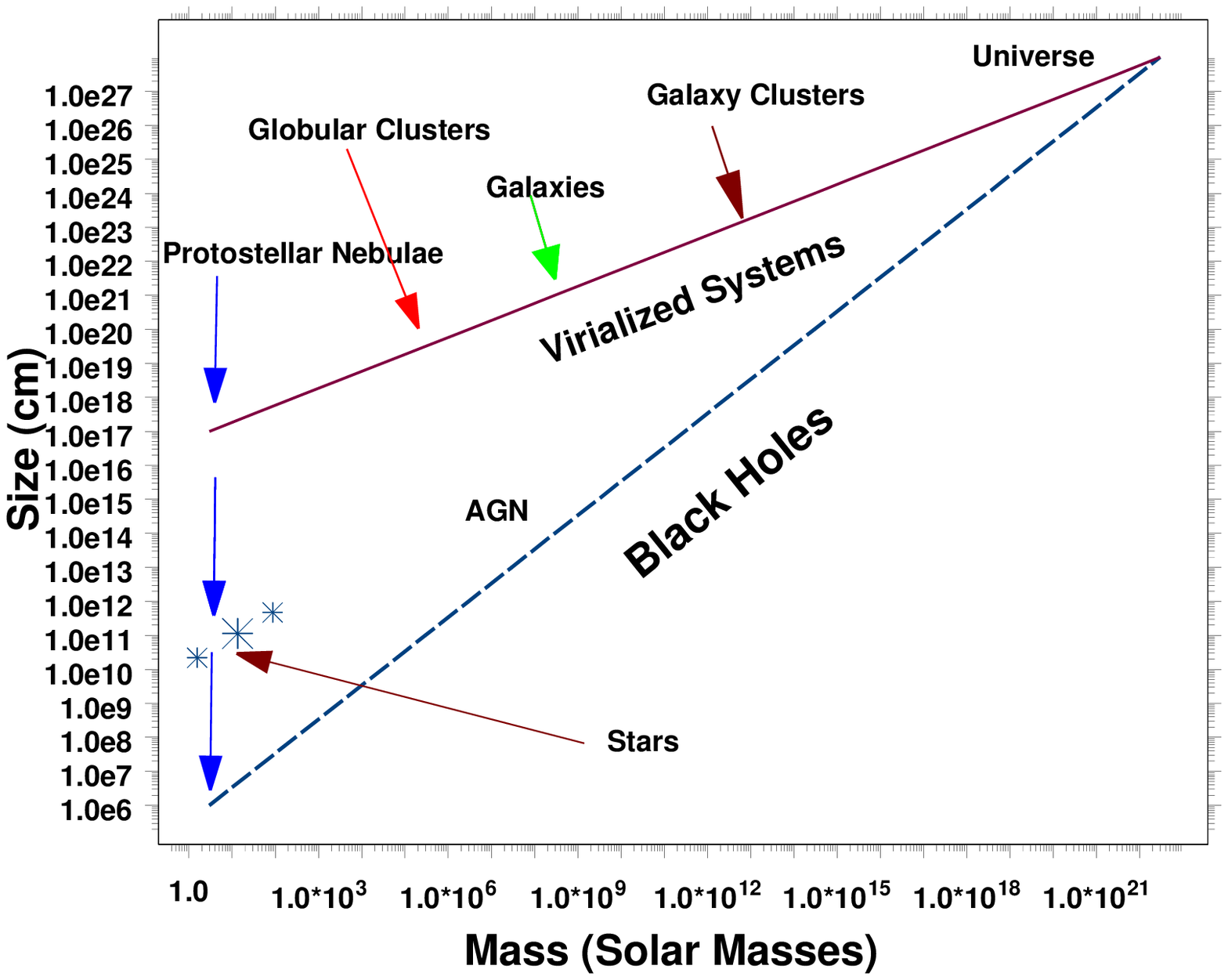}}
\hspace*{-0.1\textwidth}
\hbox{\includegraphics[width=0.68\textwidth]{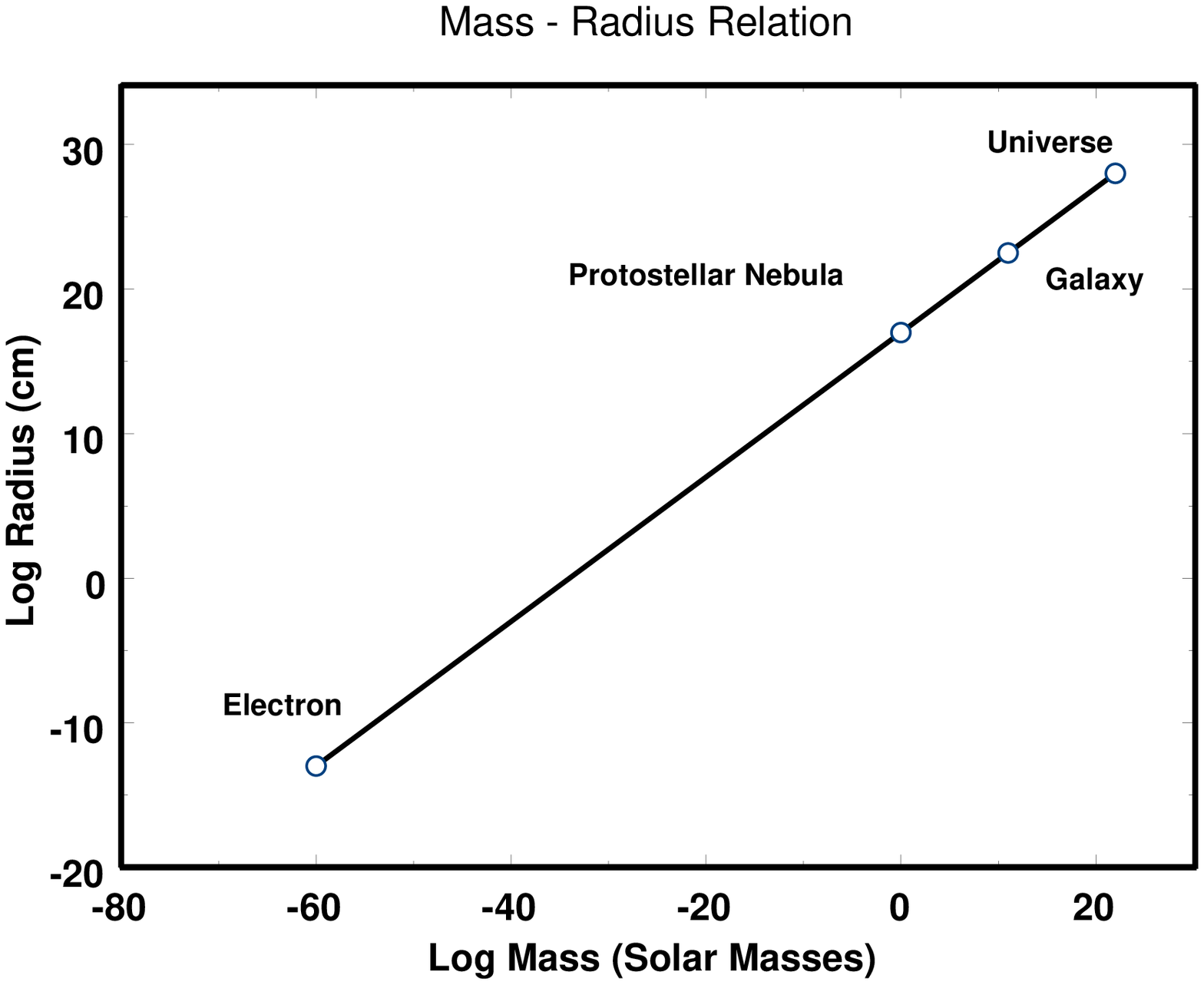}}}
\vspace*{-0.3cm} \caption{\small Left: The Mass-Radius relation $r^2
= 2M R_H$ (solid) along with black hole line $r= 2M$; arrows point
to the regions occupied by classes of the objects noted. Right: The
same relation extrapolated to the mass of the electron.}
\vspace*{-5mm}
\end{figure}

On the other hand, the success of the `slow-roll' Inflation in
confronting the CMB fluctuations has lent the confidence to venture
into notions removed from the constraints of observations such as
eternal inflation, i.e. the creation, through unlikely but
sufficiently large fluctuations, of domains (`baby' universes) that
inflate much faster than the parent domains which in their turn also
self-reproduce and so on (see e.g. \cite{Linde02} and references
therein). Each such region becomes, then, a universe of its own,
with (possibly) different values of the inflaton field $\phi$ and
possibly different values of the physical constants. With the
apparent proliferation of `universes', the question is whether our
accessible to observation domain  is special. To the best of my
understanding, a seriously considered (and perhaps prevailing) view
is that we live in the domain with the proper parameters to foster
life, thereby enunciating a truly cosmic version of the Copernican
view. So, while the inflationary scenario draws support from its
consistency with the CMB observations, some of its other (more far
reaching) consequences lie outside the domain of the observable. The
question that arises, then, is whether one should accept all these
implications as true or should consider the inflationary scenario as
an {\em ansatz} that simply provides a framework within which one
can work out and fit the CMB fluctuation data, much in the same way
that the Bohr quantum theory  did provide a resolution to the issue
of the atomic spectra. The answers to these questions lie possibly
in future more accurate observations or alternative theoretical
developments.

To provide an example of a theory that addresses coincidences with
fine tuning akin in precision to that of the standard cosmology, I
will refer to the locally scale invariant theory of gravity
considered in \cite{MK90, MK91}. This theory is defined by the
unique action ($C^{\alpha \beta \gamma \delta}$ is the Weyl tensor)
\begin{equation}
I_W =-\alpha \int d^4 x (-g)^{1/2} C^{\alpha \beta \gamma \delta}
C_{\alpha \beta \gamma \delta}\label{W2}
\end{equation}
whose static spherically symmetric geometry with a charge $Q$
reads
\begin{equation}
g_{00}=1/g_{rr} = 1 - 3 \beta \gamma - \beta(2 - 3 \beta \gamma)/r -
Q^2/(8 \alpha \gamma r) + \gamma r - k r^2 \label{MK}
\end{equation}
where $\beta,~\gamma,~k$ are integration constants. One should note
first that in this theory charge modifies geometry the same way as
mass, possibly evading the problems that the $Q^2/r^2$ term of the
Einstein gravity solution entails! For $\gamma=0, ~Q=0$ this metric
is that of Schwarzschild - de Sitter. However the linear term
(analogous to the quark potential) is totally novel and being
asymptotically non-flat, it is reasonable to associate $\gamma$ with
the inverse Hubble length $R_H$. The presence of this term provides
a {\em first principles} characteristic acceleration $2M/r^2 \simeq
1/R_H$ and suggests deviations of order 1 from the Newtonian
potential at distances such that $r^2 \simeq 2M R_H$ i.e. at a
radius that is the geometric mean of the Schwarzschild and the
Hubble radius. It is interesting to note that several classes of
virialized objects (including the Universe for which $2M \simeq
R_H$) lie on this line (Fig 2a). This is relevant to inflation
because the mass-radius relation associated with galaxies and their
clusters presumably originates in the inflationary perturbations.
One could suggest that these systematics (known as the Tully-Fisher
and Larson relations in galaxies and star forming regions
respectively) are due to the non-linear dynamics of clustering.
However, extrapolation of this relation by 60(!) orders of magnitude
to the mass of the electron (Fig. 2b), yields for the radius the
classical electron radius!!! So, it is not only the size of the
Universe at the Planck time that presents us with a fine tuning
problem. There exist numerical relations equally astounding but very
little understood even outside the standard gravity and cosmology,
apparently related to the metric of Eq. (\ref{MK}).

As suggested in \S 2 the origin of the Horizon and Flatness problems
can be traced in part to the presence of a scale in the
gravitational Lagrangian. Actually, in some Inflation variants the
universe at creation had a size equal to the Planck length and mass
equal to the Planck mass, gaining mass as it inflated. As described
above, inflation as of today does not provide estimates of the real
size of the universe, since we cannot predict how long this period
lasts and, even worse (depending on one's view), it allows for the
possibility of a huge number of disjoint domains (Multiverse). At
this section of speculations I would like to venture to a totally
different point of view which at present provides only hints on
directions that may be followed in the future. It involves the
notion of information, which, as it has been suggested, may lie at
the root of all physics \cite{lloyd}. To be sure, it is easy to see
that the Special Theory of Relativity rests on and can be formulated
on the condition of a finite, maximum information propagation speed,
namely $c$. Pursuing a similar line of thought, I would suggest
Quantum Mechanics as the framework for imposing a finite, maximum
information density, namely $h$. Within this same framework, then,
gravity appears to be the source of free energy necessary to process
the available information; as such, it also provides a sense for the
direction of time, in fact gives rise to time itself. What about the
total amount of information? A (pre-Inflation) universe of size
equal to $l_P$ and mass $M_P$ contains only one bit of information.
In my view such a universe is rather uninteresting and likely to
remain virtual. This immediately raises the issue of whether any
amount of information can be converted from virtual to real or
whether a minimum amount is necessary (while these considerations
border the metaphysical, so are those pondering the existence of
other universes totally inaccessible to us). Perhaps this is
possible only for sufficiently large number of bits [$10^{90}$?
(i.e. the number of bits the observable universe contained at the
Planck time assuming one bit per horizon); could this be {\it the}
reason gravity is so weak?) and perhaps a very specific geometrical
arrangement is needed (as discussed in \cite{Penrose1}) for their
conversion from virtual to real, presumably by the influence gravity
\cite{Penrose1}.
Such a proposal would resolve in a different way the issues of
Horizon and Flatness (but it would also need to provide the dynamics
necessary to to produce the CMB fluctuations, as discussed above).
Under the same proposal, it would be natural to also consider a
finite amount of the total available information and therefore a
closed universe. Considering that such theory, like that of Eq.
(\ref{W2}), may involve a Lagrangian with a dimensionless coupling
constant (and therefore lacking the Planck density $\Lambda_P\propto
1/l_P^4$ as its characteristic density), the present need for a
non-zero cosmological constant, some 120 orders of magnitude smaller
than $\Lambda_P$, should be considered with some caution.

{\bf Acknowledgments} This talk was presented as a public lecture at
the Academy of Athens as part of the ``Chaos in Astronomy 2007"
symposium. I would like to thank the organizing committe for the
invitation, as well as for their financial support.

\end{document}